\documentclass[prd,showpacs,preprintnumbers,amsmath,amssymb,nofootinbib,superscriptaddress,english]{revtex4}
\usepackage{graphicx}
\usepackage{dcolumn}
\usepackage{bm}
\usepackage{epsfig}
\usepackage{graphicx}
\usepackage{hyperref}
\usepackage[usenames]{color}
\usepackage{url}

\newcommand{\remove}[1]{}

\newcommand{\dd}{\mathrm{d}} % For the derivatives
\newcommand{\rr}{\mathrm}

\newcommand{\Tbr}{T_{\rr b}}
\newcommand{\Tspin}{T_{\rr{s}}}
\newcommand{\Tg}{T_{\rr g}}
\newcommand{\deltaB}{\delta_{\rr b}}
\newcommand{\deltaDM}{\delta_{\rr{c}}}
\newcommand{\deltaTg}{\delta_{\rr T}}
\newcommand{\omb}{\Omega_{\rr b}}

\newcommand{\xe}{x_{\rr e}}
\newcommand{\Nant}{N_{\rr A}}
\newcommand{\Dmax}{D_{\rr{max}}}
\newcommand{\mum}{\mu_{\rr m}}
\newcommand{\sigmaT}{\sigma_{\rr T}}
\newcommand{\Sgauge}{S_{\rr{gauge}}}

\newcommand{\Mpc}{\rr{Mpc}}

\newcommand{\CAMB}{\texttt{CAMB}}
\newcommand{\RECFAST}{\texttt{RECFAST}}

\def\be{\begin{equation}}
\def\ee{\end{equation}}
\def\ba{\begin{eqnarray}}
\def\ea{\end{eqnarray}}

\hypersetup{
    colorlinks=true,
    linkcolor=green,
    citecolor=blue,
}

\begin{document}

\title{Signatures of Modified Gravity on the 21-cm Power Spectrum at Reionisation}

\author{Philippe~Brax}
\email[Email address: ]{philippe.brax@cea.fr}
\affiliation{Institut de Physique Th\'eorique, CEA, IPhT, CNRS, URA 2306, F-91191Gif/Yvette Cedex, France}

\author{S\'ebastien~Clesse}
\email[Email address: ]{s.clesse@damtp.cam.ac.uk}
\affiliation{DAMTP, Centre for Mathematical Sciences, University of Cambridge, Wilberforce Road, Cambridge CB3 0WA, UK}

\author{Anne-Christine~Davis}
\email[Email address: ]{a.c.davis@damtp.cam.ac.uk}
\affiliation{DAMTP, Centre for Mathematical Sciences, University of Cambridge, Wilberforce Road, Cambridge CB3 0WA, UK}

\date{\today}

\begin{abstract}
Scalar modifications of gravity have an impact on the growth of structure. Baryon and Cold Dark Matter (CDM) perturbations grow anomalously
for scales within the Compton wavelength of the scalar field. In the late time Universe when reionisation occurs, the spectrum of the 21cm brightness temperature is thus affected. We study this effect for chameleon-f(R) models, dilatons and symmetrons.  Although the $f(R)$ models are more tightly constrained by solar system bounds, and effects on dilaton models are negligible, we find that symmetrons where the phase transition occurs before $z_{\star} \sim 12$ will be detectable for a scalar field range as low as $5\ {\rm kpc}$. For all these models, the detection prospects of modified gravity effects are higher when considering modes parallel to the line of sight where very small scales can be probed. The study of the 21 cm spectrum thus offers a complementary approach to testing modified gravity with large scale structure surveys. Short scales, which would be highly non-linear in the very late time Universe when structure forms and where modified gravity effects are screened, appear in the linear spectrum of 21 cm physics, hence deviating from General Relativity in a maximal way.
\end{abstract}

\maketitle

\section{Introduction}

A major challenge  for theoretical cosmology  is the explanation of the recent acceleration of the Universe's expansion~\cite{Riess:1998cb}.   In the standard $\Lambda$-CDM scenario, it is the consequence of the existence of a cosmological constant, although it could be also due to a dark energy fluid whose origin has yet to be determined~\cite{Copeland:2006wr}.
Models of modified gravity~\cite{Clifton:2011jh} complement dark energy scenarios and provide an explanation of the absence of long range fifth force effects in the solar system and laboratory experiments. Indeed  most involve at least one scalar field coupled to matter, and eventually an environmental dependence leading  to a screening mechanism of the scalar field in high density regions~\cite{Khoury:2010xi}.   This mechanism is  an essential ingredient for the models to pass the stringent constraints on the possible modifications of gravity in the laboratory~\cite{Adelberger:2002ic},  the solar system~\cite{1990MNRAS.247..510S}, and the galactic environments~\cite{Pourhasan:2011sm}.  Moreover,  the scalar fields are required to sit at the minimum of the density dependent effective potential prior to Big Bang Nucleosynthesis (BBN), so that catastrophic modifications in the formation of light elements are avoided.

Numerous models of this type have been proposed.  Let us mention the chameleons~\cite{Khoury:2003rn,Mota:2006fz,Brax:2004px,Brax:2005ew,Brax:2004ym,Brax:2010kv,Gannouji:2010fc,Hees:2011mu}, involving a thin shell shielding the scalar field in dense bodies, the symmetrons~\cite{Pietroni:2005pv,Olive:2007aj,Hinterbichler:2010es,Hinterbichler:2011ca,Brax:2011pk,Davis:2011pj,Clampitt:2011mx}, involving a symmetry breaking potential so that the scalar field is decoupled from matter at high densities, the dilatons~\cite{Brax:2010gi,Brax:2011ja}, where the coupling to gravity turns off in dense environments, and the $f(R)$ models~\cite{Starobinsky:1980te,Carroll:2003wy,Carroll:2004de,Faulkner:2006ub,Navarro:2006mw,Amendola:2006we,Carloni:2007br,Song:2006ej,Li:2007xn,Sawicki:2006jj,Brax:2011ja}, which are a sub class of chameleon models~\cite{Brax:2011ja}.

At the homogeneous level, all these scenarios coincide with the $\Lambda$-CDM model.   However, the evolution of linear perturbations differ.  As a consequence, models of modified gravity can induce observable signatures in the matter power spectrum at redshifts $z \lesssim 2$, and less importantly in the cosmic microwave background~\cite{Hojjati:2011ix,Brax:2011ja} at $z \simeq 1100$.   Modified gravity models can also be probed with weak lensing (see e.g.~\cite{Schmidt:2008hc}).  During the dark ages and the reionisation period, \textit{i.e.} in the range $1100 > z \gtrsim 6 $, no cosmological signal have yet been observed.   Such observations would be however of great interest for cosmology,  especially for the study of modified gravity through the time evolution of the matter perturbations.

In the near future, this gap is expected to be partially filled with the observation of the 21cm signal from  reionisation~\cite{Madau:1996cs,Ciardi:2003hg,Loeb:2003ya,Furlanetto:2006jb,Pritchard:2011xb}, and maybe in the more distant future, from the dark ages.   During  reionisation, transitions between the fundamental hyperfine levels of neutral hydrogen atoms  are possible, via the Wouthuysen-Field effect involving the absorption and re-emission of Lyman-$\alpha$ photons from the first stars (for a review, see Ref.~\cite{Furlanetto:2006jb}).  These induce the 21cm signal corresponding to a stimulated emission of 21cm photons against the Cosmic Microwave Background (CMB) radiation.  The 3D power spectrum of the 21cm radiation maps the baryon distribution and thus  is sensitive to modifications of gravity. In this paper, we explore for the first time the effects of modified gravity on the 21cm power spectrum at reionisation, and discuss the detectability of such effects with instruments of  future generation  giant Fast Fourier Transform radio-telescopes~\cite{Tegmark:2008au,Tegmark:2009kv}.

Several parameterisations of modified gravity have been proposed~\cite{Bertschinger:2006aw,Hu:2007pj,Jain:2007yk,Amendola:2007rr,Zuntz:2011aq,Bertschinger:2008zb,Song:2008vm,Bean:2010zq,Daniel:2010ky,Pogosian:2010tj,Zhao:2011te,Skordis:2008vt,Ferreira:2010sz} for the evolution of linear perturbations such as, for instance, through the Poisson equation, $- k^2 \Phi  = 4 \Pi (1 + \nu ) G a^2 \delta \rho_{\rr m}$ and $ \Psi = (1+ \gamma) \Phi$, where $\delta \rho_{\rr m}$ is the matter density perturbation and where $\Phi$ and $\Psi$ are the potentials in the Newtonian gauge.  This parametrization involves two functions $\nu(k,a)$ and $\gamma(k,a)$ that depend both on time and on the perturbation wavenumber $k$.  For $f(R)$ models, those functions depend on  $B = (f_{RR} / f_R ) H \dd R / \dd H $ together with $f_{R0} $ today~\cite{Song:2006ej}.
In this paper, we adopt the parameterisation proposed in Refs.~\cite{Brax:2011aw,Brax:2012gr}, for which the  action and dynamics can be fully and uniquely reconstructed from two time-dependent functions: the coupling to matter $\beta(a)$ and the scalar field mass $m(a)$. Of course, in the $f(R)$ case, it coincides with the usual approach. Moreover, for all these models, the $\nu$ and $\gamma$ functions can be explicitly obtained as a function of $m(a)$ and $\beta(a)$.

Using the $(m(a),\beta(a))$ parameterisation, we calculate the signatures of $f(R)$, dilaton, chameleon and symmetron models on the 21cm power spectrum at reionisation, as well as on the present matter power spectrum in the linear approximation.   For each model, we discuss the range of parameters that could be probed by 21cm experiments and compare it to the constraints from local experiments.   Finally, we evaluate how the coupling to photons could be bounded via 21-cm constraints on the variation of the fine structure constant $\alpha$.
In all cases we find that the 21 cm signal obtained by varying the modes parallel to the line of sight is the most relevant. In the $f(R)$ case, we expect the constraints to be less stringent than the ones from the gravity tests in the solar system. For dilatons, the signal is found to be negligible while symmetron models with a transition at a redshift larger than $z_{\star} \sim 12$ could be detected even when the range of the symmetron interaction now is as low as $5$ kpc.

This paper is organised as follows:  In Sec.~\ref{sec:21cm}, we briefly summarise the physics of the 21cm signal at the period of reionisation and derive its 3D power spectrum.   We also give the evolution of the baryon and dark matter perturbations after the time of last scattering for the $\Lambda$-CDM model.  In Sec.~\ref{sec:MGmodels}, we show  how these equations are modified in the context of scalar models and describe the dynamics of $f(R)$, symmetron, chameleon and dilaton models using the reconstruction of the scalar field dynamics from the parametrization of Ref~\cite{Brax:2011aw,Brax:2012gr}.  In Sec.~\ref{sec:21cmeffects} we give the specifications of the considered FFTT radio-telescope as well as the forecast errors on the 21cm power spectrum for single redshift measurements.    We then evaluate the range of parameter values leading to observable effects and compare to the constraints from local tests of gravity.   Our results are summarised in the conclusion.

%These data, combined with galaxy, solar system, and laboratory experiments,

\section{21cm signal from reionisation} \label{sec:21cm}

In this section, we briefly review the basics of the 21cm cosmic background from the period of reionisation and  refer to Refs.~\cite{Furlanetto:2006jb,Lewis:2007kz,Pritchard:2011xb,Mao:2008ug} for a more exhaustive description  of the signal and its relevance to cosmology.  The 21cm signal corresponds to an absorption (or a stimulated emission) of 21cm photons from (against) the CMB, induced by the transitions between the hyperfine ground state of neutral hydrogen (HI) atoms. During the dark ages, after the thermal decoupling of the baryon gas at $z \sim 200$, the hyperfine level population is shifted away from thermal equilibrium with photons because of the spin changing collisions between HI atoms, and between HI and free electrons.   In this paper, we focus on the 21cm signal from the period of reionisation ($z~\sim 10$).  During this period, collisions are rarefied and hyperfine transitions are driven by the so-called Wouthuysen-Field effect, via the absorption and the re-emission of Lyman-$\alpha$ photons coming from the first  luminous objects that reionise the Universe.

\subsection{21cm Power Spectrum}

At the homogeneous level, the difference between the observed brightness temperature and the CMB temperature at a given observed energy $E$ is given by~\cite{Lewis:2007kz}
\begin{equation}
T_{\rr B} (E) = \left( 1 - \rr e ^{- \tau_E} \right) \left(\frac{\Tspin - T_\gamma }{1+z} \right)_{a_E} ~,
\end{equation}
where
\begin{equation}
\tau_E = \frac{3 c^3  n_{\rr{HI}}(a_{E})  A_{10} h_{\rr p} a_E^2} {32 \pi
  \Tspin(a_E) k_B \nu_{21}^2 (\dd v_\parallel / \dd r)}~,
\end{equation}
is the 21cm optical depth.   $\nu_{21}$ is the frequency of the 21 cm line in the rest frame.   $a_E=1/(1+z_E)$ is the scale factor for which $E = a_E
E_{21}$ with $E_{21}=h_p \nu_{21}$, the energy of the 21 cm spin flip
transition.  $A_{\rr{10} } = 2 \pi \alpha \nu_{\rr{21}}^3 h_{\rr p}^2 / (3 c^4 m_{\rr e} ^2 )  \simeq 2.869 \times 10^{-15}
\rr{s^{-1}}$  is the Einstein coefficient of spontaneous emission. $ \dd v_\parallel / \dd r$ is the gradient of the physical velocity along the line of sight and $r$ is the comoving distance.  At the homogeneous level, there is no peculiar velocity and thus $ \dd v_\parallel / \dd r = a_E H(a_E)$.
$n_{\rr{HI}}$ is the number density of neutral hydrogen atoms.
It is related to the baryon
number density $n_{\rr{b}}$ and the neutral fraction $x_{\rr H}$ via
 $n_{\rr HI}= x_{\rr H}n_{\rr{b}}(1-f_{\rr{He}})$, where $f_{\rr{He}}$  is the Helium fraction.

During the reionisation process, $\tau_E \ll 1$, so that the 21 cm
brightness temperature is well approximated by
\begin{equation} \label{eq:Tbrionexact}
T_{\rr B} (E  ) =  \frac{3 c^3  x_{\rr H}(a_E) n_{\rr{b}}(a_E) (1-f_{\rr{He}}) A_{10} h_{\rr p} }{32
  \pi k_B  \nu_{21}^2 H(a_{ E})}  \left. \frac{\Tspin - T_\gamma }{ \Tspin
(1+z)} \right|_{a_E}~.
 \end{equation}

After perturbing at the linear level the baryon density, the ionized fraction and the gradient of the radial velocity, the
brightness temperature  in a
given direction ${\bf e}$ is given by
\begin{equation}   \label{eq:T_B_reion}
T_{\rr B} (E,{\bf e} ) = \frac{ \bar T_{\rr B} }{\bar x_{\rr H}} \left[  1 - \bar x_{\rr i} ( 1 + \delta_{\rr i} ) \right] ( 1 + \delta _{\rr b}  ) \left( 1 - \delta_{\rr v} \right) ~,
\end{equation}
where a bar denotes homogeneous quantities. $\delta  _{\rr b}$ and $ \delta_{ i}$ are respectively the
relative baryon and ionized fraction perturbations, and
\begin{equation}
\delta_{\rr v} \equiv \frac{1}{aH}  \frac{\partial v_\parallel}{ \partial r} ~
\end{equation}
is the radial gradient of baryon peculiar velocities.  Let us note that in order to obtain Eq.~\ref{eq:T_B_reion}, we have followed Ref.~\cite{Mao:2008ug} and assumed that there exists a redshift range during the reionisation with $\Tspin \gg
T_\gamma$ so that $(\Tspin - T_\gamma )/\Tspin\simeq 1 $.   With this assumption, the fluctuations of the spin and photon temperatures are typically second order and can be conveniently neglected.

Defining by $\theta=\partial_i v^i$ the divergence of the velocity field we have
\be
\frac{\partial v^r}{\partial r}=\frac{\mathbf {\hat n.k}}{k} \theta_b~,
\ee
where $\mathbf{\hat n}$ is the line of sight vector and $\mathbf{k}$ the wave vector of a linear perturbation.
The perturbed conservation of matter gives
\be
\theta_b=-\delta_b'
\ee
in conformal time.
In the long time regime, the baryon contrast grows with a growing mode $D_+$ which is not the scale factor $a$ anymore in modified gravity, hence
\be
\delta_b= D_+ \delta_{b0}~,
\ee
from which we get
\be
\left( \frac{\dd \eta }{\dd \ln D_+} \right) \theta= -\delta_b~,
\ee
and therefore
\be
\delta_{\rr v}=- \frac{1}{aH} \left(\frac{\dd \ln D_+}{\dd \eta} \right) \tilde \mu^2 \delta_b~,
\ee
 where $\tilde \mu \equiv \mathbf{k \cdot \hat{\bf n }}/k
 $  is the cosine of the angle between the line of sight and the wave vector  $\mathbf{k} $.
 The first factor $(\dd \ln D_+)/(aH \dd \eta)$ is equal to one in the absence of modified gravity. In the following we will use
 \be \label{eq:mu}
 \mu^2 \equiv \left(\frac{\dd \ln D_+}{aH \dd \eta}\right)\left(\frac{\mathbf{\hat n.k}}{k}\right)^2~.
 \ee
From Eq.~(\ref{eq:T_B_reion}),  the brightness
temperature perturbation in  Fourier space is given by
\begin{eqnarray}
 \Delta T_{\rr B} (\mathbf k)  & = &  \frac{ \bar T_{\rr B} }{\bar
  x_{\rr H}} \left\{ (1+\mu^2) \delta_{\rr b} -
 \bar x_{\rr i} \left[ \delta_{\rr i}  - (1+\mu^2)
 \delta_{\rr b}  \right] \right\} ~.
\label{eq:DeltaTb}
\end{eqnarray}
Therefore the 21cm 3D-power spectrum reads~\cite{Mao:2008ug}
%\begin{widetext}
\be   \label{eq:PTb}
P_{\Delta T_{\rr B}} (\mathbf k )   =  \left( \frac { \bar T_{\rr B}
}{\bar x_{\rr H}} \right)^2  \left\{  \left[  \bar
  x_{\rr H}^2 P_{\rr{bb}}(k) - 2 \bar x_{\rr H} \bar x_{\rr i}
  P_{\rr{ib}}(k) +  \bar x_{\rr i}^2 P_{\rr{ii}} (k)\right]
  +  2 \mu^2 \left[ \bar x_{\rr H}^2 P_{bb}(k ) -   \bar x_{\rr H} \bar x_{\rr i} P_{\rr{ib}}( k)  \right]
   +   \mu^4 \bar x_{\rr H}^2 P_{\rr{bb}}(k ) \right\} ~,
\ee
%\begin{eqnarray}  \label{eq:PTb}
%P_{\Delta T_{\rr B}} (\mathbf k )  & = & \left( \frac { \bar T_{\rr B}
%}{\bar x_{\rr H}} \right)^2  \left\{  \left[  \bar
 % x_{\rr H}^2 P_{\rr{bb}}(k) - 2 \bar x_{\rr H} \bar x_{\rr i}
 % P_{\rr{ib}}(k) +  \bar x_{\rr i}^2 P_{\rr{ii}} (k)\right] \right. \nonumber \\
 % & + & 2 \mu^2 \left[ \bar x_{\rr H}^2 P_{bb}(k ) -   \bar x_{\rr H} \bar x_{\rr i} P_{\rr{ib}}( k)  \right] %\nonumber \\
%  &  + & \left.
%\mu^4 \bar x_{\rr H}^2 P_{\rr{bb}}(k ) \right\} ~,
%\end{eqnarray}
%\end{widetext}
where the power spectra $P_{\alpha\beta} (k)$ with $\alpha, \beta=
{\rr i, \rr b}$ are defined from  $\langle \delta_{\alpha}
(\mathbf k),  \delta_{\beta} (\mathbf k') \rangle  =  (2 \pi)^3
\delta^3 ( \mathbf{k - k'} ) P_{\alpha\beta} (k)$.

 From Eq.~(\ref{eq:PTb}), one can notice that the $\mu^4$ component does not depend on $P_{\rr{ib}}$ and $P_{\rr{ii}}$.  By measuring the 3D power spectrum, it is therefore in principle possible to separate cosmology from the astrophysical contaminants related to the reionisation process. In this paper, we will assume for simplicity an optimistic reionisation model~\cite{Mao:2008ug} for which $P_{\rr{ib}} = P_{\rr{ii}} = 0$ at the redshift of interest.
 % and we will calculate the 21cm power spectrum only for orthogonal modes.

A dedicated 21cm experiment does not directly measure the comoving modes ${\bf k}$ but the angular positions in the sky and the signal frequency.  It is therefore convenient to determine the observable 21cm power spectrum in the Fourier dual of the space of angular positions and frequencies, the so-called $\bf u$-space.  It is related to the $\bf k$-space through the relations
\begin{equation}
\mathbf u_\bot = D_{\rr A} \mathbf k_\bot~,
\end{equation}
\begin{equation}
u_{\parallel} = y k_{\parallel}~,
\end{equation}
where $D_{\rr A} $ is the comoving angular distance, given in a flat Universe by
\begin{equation}
D_{\rr A} (z) = c  \int_0 ^z \frac{1}{H(z')} \dd z'~,
\end{equation}
and where $y(z)$ is the conversion factor between comoving distances and frequency intervals,
\begin{equation}
y(z) = \frac{\lambda_{21} (1+z)^2 }{H(z)}~.
\end{equation}
The 3D power spectrum of the 21cm brightness
temperature in  $\mathbf u$-space reads
\begin{equation}
P_{\Delta T_{\rr B} } ( \mathbf u ) = \frac{P_{\Delta T_{\rr B} }
  (\mathbf k) }{D_{\rr A} ^2 y}~.
\end{equation}

\subsection{Evolution of Baryon Perturbations}

When baryon and dark matter perturbations re-enter inside the horizon, they feel their combined gravity. Baryon perturbations feel also the pressure.   Assuming a $\Lambda$-CDM model at the background level, cold dark matter and baryon perturbations evolve according to~\cite{Naoz:2005pd} in the absence of modified gravity
\be  \label{eq:deltac}
\ddot \delta_{\rr{c}} + 2 H \dot \delta_{\rr c} = \frac 3 2 H^2 \left( \omb \deltaB + \Omega_{\rr c} \deltaDM  \right)~,
\ee
\be   \label{eq:deltab}
\ddot \delta_{\rr b} + 2 H \dot \delta_{\rr b} = \frac 3 2 H^2 \left( \omb \deltaB + \Omega_{\rr c} \deltaDM  \right) - \frac{k^2}{a^2} \frac{k_{\rr B} \Tg}{\mum} \left( \deltaB + \deltaTg \right)~,
\ee
where $\Omega_{\rr b}$ is the baryonic fraction  which is a constant $f= \rho_{\rr b}/(\rho_{\rr c}+\rho_{\rr b})$ in the matter era, $\Omega_{\rr c}$ the CDM fraction, $\mum$ is the mean molecular weight and $\deltaTg$ is the gas temperature perturbation.
The homogeneous gas temperature $\Tg$  evolves during the dark ages according to~\cite{Naoz:2005pd}
\be  \label{eq:Tb}
\frac{\dd \Tg}{\dd t} = - 2 H \Tg + \frac{x_{\rr e}}{t_\gamma a^4} \left( T_\gamma - \Tg \right)~,
\ee
where
\be
t_\gamma ^{-1} \equiv \frac{8 \rho_\gamma ^0 \sigmaT c }{3 m_{\rr e}} = 8.55 \times 10^{-13} \rr{yr}^{-1}~.
\ee
The last term on the r.h.s. of Eq.~(\ref{eq:Tb}) accounts for the energy injection due to the Compton scattering between CMB photons and the residual free electrons.  During the period $1100 \gtrsim z \gtrsim 200$, the Compton heating drives $\Tg \rightarrow T_\gamma$.  After $z \sim 200 $, due to the expansion, the gas temperature decouples from the radiation and evolves as $\Tg \propto 1/a^2 $, as expected for an adiabatic non relativistic gas in expansion.  We have assumed that the non-trivial evolution of the gas temperature during the reionisation has only a negligible effect on the baryon perturbations.

Eq.~(\ref{eq:Tb}) can be perturbed at the linear level.  On small scales and assuming that photon density and photon temperature perturbations can be neglected, the gas temperature perturbations evolve according to~\cite{Naoz:2005pd}
\be \label{eq:deltaTg}
\dot \delta_{\rr T} = \frac{2}{3} \dot \delta_{\rr b} - \frac{\xe (t) }{t_\gamma a^4} \frac{T_\gamma}{\Tg} \deltaTg~.
\ee

At high redshifts, the interaction $ p + e \leftrightarrow H + \gamma$ maintains the species in equilibrium and the free electron fraction is given by the Saha equation
\begin{equation} \label{eq:saha}
\frac{x_{\rr e } ^2}{1 - x_{\rr e}} = \frac{1}{n_{\rr b}  } \left( \frac{m_{\rr e } T}{2 \pi} \right)^{3/2} \rr e^{-\epsilon_0 / T  }~,
\end{equation}
where $\epsilon_0 = m_{\rr e} + m_{\rr p} - m_{\rr H} = 13.6$ eV.
%Because $n_{\rr b } \ll n_\gamma$, when $T \sim \epsilon_{0} $, the right hand side is of the order of $10^{15}$ and the free electron fraction remains $x_{\rr e} \simeq 1$.  The recombination therefore only occurs at $T \ll \epsilon_0 $.
Near the redshift of the last scattering, the equilibrium is not maintained anymore and the Saha equation becomes inaccurate.  One therefore needs to solve the Boltzmann equation for $x_{\rr e}$.  A good approximation\footnote{For more accurate results, the recombination can be calculated numerically by using the \RECFAST \ code~\cite{website:recfast}, taking into account  additional effects like Helium recombination and a 3-level atom. } is given in Refs.~\cite{Seager:1999bc,Seager:1999km},
\begin{equation} \label{eq:recomb}
\frac{\dd x_{\rr e}}{\dd t} = \left[ (1-x_{\rr e} ) \beta_{\rr i} - x_{\rr e}^2 n_{\rr b} \alpha^{(2)} \right]~,
\end{equation}
where
\begin{equation}
\beta_{\rr i} \equiv \alpha^{(2)} \left( \frac{m_{\rr e} T }{2 \pi} \right)^{3/2} \rr e^{- \epsilon_{\rr 0} / T}
\end{equation}
is the ionization rate, and where
\begin{equation}
\alpha^{(2)} = 9.78 \frac{\alpha^2}{m_{\rr e}^2} \left( \frac{\epsilon_0}{T} \right)^{1/2} \ln \left(  \frac{\epsilon_0}{T} \right)
\end{equation}
is the recombination rate.  The superscript $^{(2)} $ indicates that recombination in the ground state is not relevant.  Indeed, this process leads to the production of a photon that ionizes immediately another neutral atom and thus there is no net effect.  For the purpose of this work, we have calculated the free electron fraction with Eq.~(\ref{eq:saha}) for $x_{\rr e} > 0.99$, and with Eq.~(\ref{eq:recomb}) for  $x_{\rr e} < 0.99$.

Equations (\ref{eq:deltac}), (\ref{eq:deltab}), (\ref{eq:Tb}), (\ref{eq:recomb}) and (\ref{eq:deltaTg}) form a closed set of equations.  In order to calculate the 21cm power spectrum, these have been integrated numerically.  Initial conditions for the baryon, cold dark matter and gas temperature perturbations are provided at a redshift $z=950$ by the \CAMB \   code~\cite{Lewis:1999bs}.   In the next section, the equivalent equations to  Eqs.~(\ref{eq:deltac}) and (\ref{eq:deltab}) for modified gravity models will be described.   We have only considered scenarios for which modified gravity effects at $z>950$ are negligible.

\section{Modified Gravity Models} \label{sec:MGmodels}

\subsection{Scalar Field and Modified Gravity }

Scalar-tensor theories are characterised by their coupling to
matter and their interaction potential,
\begin{eqnarray}
S = \int \sqrt{-g} \dd^4 x\left[\frac{R}{2\kappa} -
\frac{1}{2}(\nabla \phi)^2 - V(\phi)\right] + S_{\rm m},
\end{eqnarray}
where $\phi$ is the scalar field and $V(\phi)$ its potential,
$S_{\rm m}\equiv S_{\rm m}\left[\Psi^{i}, \tilde{g}_{\mu
\nu}\right]$ the matter action with $\Psi^{i}$ the matter fields
which are \emph{minimally} coupled to the Jordan frame metric
$\tilde{g}_{\mu \nu} =  A^2(\phi)g_{\mu\nu}$; $g_{\mu\nu}$ is the
Einstein frame metric, which is used to compute the Ricci scalar
$R$; $\kappa_4^2=\kappa=8\pi G=m_{\rm pl}^{-2}$ where $G$ is Newton's constant and
$m_{\rm pl}$ the reduced Planck mass.

The field equations are obtained by varying the action $S$ with
respect to the field $\phi$, and we have
\begin{eqnarray}
\square\phi &=& V_{,\phi}(\phi)-A_{,\phi}(\phi)A^3(\phi)\tilde{T},\nonumber\\
\tilde{T} &=& \tilde{g}_{\mu\nu}\tilde{T}^{\mu\nu},\\
\tilde{T}^{\mu\nu} &=& \frac{2}{\sqrt{-\tilde{g}}}\frac{\delta
S_{\rm m}}{\delta\tilde{g}_{\mu\nu}}~,
\end{eqnarray}
where we have defined the Jordan frame energy momentum tensor
$\tilde T_{\mu\nu}$ that is related to the Einstein frame one by
$T^{\mu}{}_{\nu} = A^3(\phi) \tilde{T}^{\mu \rho}\tilde{g}_{\rho
\nu}$. The Klein-Gordon  equation for $\phi$ and the Einstein
equations  become
\begin{eqnarray}
\square\phi &=& V_{\mathrm{eff},\phi}(\phi,T)~,\\
R^{\mu \nu} - \frac{1}{2}R g^{\mu \nu} &=& \kappa T^{\mu \nu}_{\rm tot}~, \\
\end{eqnarray}
in which the scalar field is governed by an effective potential
\begin{eqnarray}
V_{{\rm eff}}(\phi,T) &\equiv& V(\phi)-A(\phi)T~,
\end{eqnarray}
and the total energy momentum is given by
\begin{eqnarray}
T^{\mu \nu}_{\rm tot} = A(\phi) T^{\mu \nu}  - g^{\mu \nu} V(\phi)
+ \nabla^{\mu}\phi \nabla^{\nu}\phi - \frac{1}{2}g^{\mu \nu}
(\nabla \phi)^2~,
\end{eqnarray}
which satisfies the following conservation equation
\begin{eqnarray}
\nabla_{\mu}T^{\mu\nu} &=& \frac{A_{,\phi}}{A}\left(Tg^{\mu\nu} -
T^{\mu\nu}\right)\nabla_{\mu}\phi~.
\end{eqnarray}
Note that this implies that for pressureless matter with
$T^{\mu\nu} = \rho_m u^{\mu}u^{\nu}$ and  $u^{\mu}$ the  4-velocity
($u^{\mu}u_{\nu} = -1$), we have $-T = \rho_{\rm m}$ and
$\rho_{\rm m}$ is conserved independently of $\phi$:
\begin{eqnarray}
\nabla_{\mu}(\rho_{\rm m}u^{\mu}) = 0~,
\end{eqnarray}
or equivalently
\be
\dot \rho_m +3h \rho_m=0~,
\ee
where $\dot\rho= u^\mu \nabla_{\mu} \rho$ and $3h=\nabla_\mu u^\mu$. The continuity equation is
\be
\dot u^\mu + \kappa_4 \beta \dot \phi u^\mu= - \kappa_4 \beta \partial^\mu \phi~,
\ee
where  the coupling to matter is defined to be:
\be
\beta= m_{\rm Pl} \partial_\phi \ln A~.
\ee
We are interested in models of modified gravity with screening properties in dense environments due to the non-linearity of the interaction potential
$V$ and the coupling function $A$. Examples of such models are chameleons, dilatons and symmetrons. f(R) models are chameleons in disguise written in the Jordan frame.

\subsection{Perturbations}

\begin{figure}[h!]
\begin{center}
\includegraphics[height=8cm]{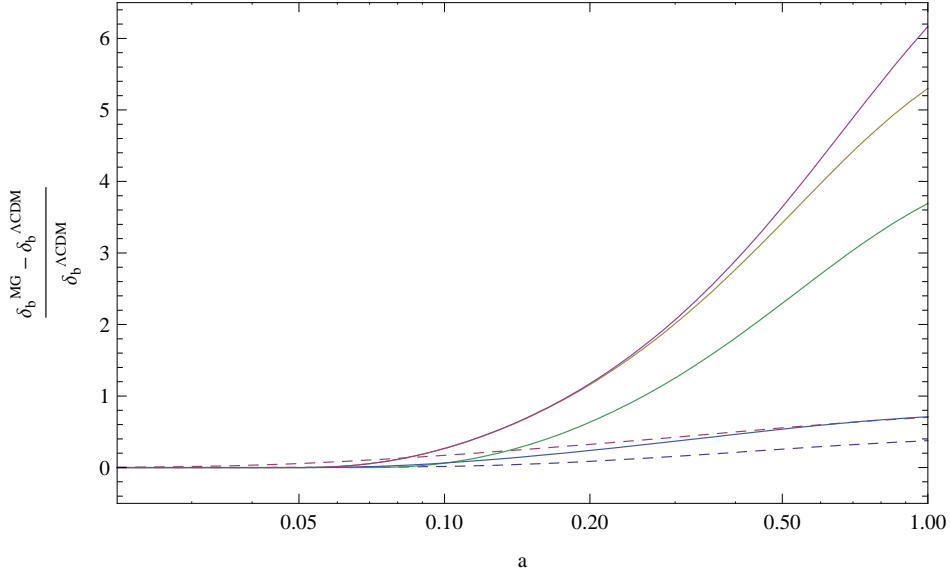}
\caption{Relative differences between the linear evolution of baryon perturbations with $k=0.1 ~\Mpc^{-1}$ for the $\Lambda$-CDM model and for $f(R)$ and symmetron models.  Plain curves are for the symmetron model.  From top to bottom lines, the model parameters are respectively $ m_0 = 0.01~\Mpc^{-1}, \beta_0 =1, z_\star = 20$ (red curve),   $ m_0 = 0.1Ê~\Mpc^{-1}, \beta_0 =1, z_\star = 20$ (yellow curve), $ m_0 = 0.1~\Mpc^{-1}, \beta_0 =1, z_\star = 14 $ (green cuvre) and $ m_0 = 0.1 ~\Mpc^{-1}, \beta_0 =0.5, z_\star = 20$ (blue curve).  Dashed curves are for the $f(R)$ model, with $m_0 = 10^{-4} \Mpc^{-1}$ (top red dashed curve) and $m_0 = 10^{-3} \Mpc^{-1}$ (bottom blue dashed curve).
%The numerical integration of Eqs. starts from z=950 with initial conditions provided by the numerical code CAMB
}
\label{fig:baryonevol}
\end{center}
\end{figure}

We will focus on models where gravity is only modified at late times well after last scattering.
The perturbation equations follow from the conservation of matter and continuity equations.
After last scattering, baryons and photons decouple while the baryons and CDM start evolving in a coupled way under the influence of gravity. The conservation equations become
\be
\delta_{\rr b}'=-\theta_{\rr b}
\ee
and
\be
\delta_{\rr c}'=-\theta_{\rr c}~,
\ee
while the continuity equations are
\be
\theta_{\rr b}'+{\cal H} \theta_{\rr b}= k^2 \Phi + \beta_{\rr b} k^2 \kappa_4 \delta\phi + k^2 c_{\rr b}^2\delta_{\rr b}~,
\ee
where we are taking into account the baryonic speed of sound $c_{\rr b}$, and for CDM
\be
\theta_{\rr c}'+{\cal H} \theta_{\rr c}= k^2 \Phi + \beta_{\rr c} k^2 \kappa_4 \delta\phi~.
\ee
We have defined the divergence of the velocity fluid for both fluids $\theta= \partial^i v_i$
and generalised the models by considering that baryons and CDM couple differently to the scalar field.
The Poisson equation is now
\be
\Phi=- \frac{3}{2k^2} {\cal H}^2 (\Omega_{\rr b}\delta_{\rr b} +\Omega_{\rr c} \delta_{\rr c})~.
\ee
Similarly the
scalar field satisfies the Klein-Gordon equation and in the sub-horizon limit becomes
\be
\kappa_4 \delta \phi= -  3{\cal H}^2 \frac{\beta_{\rr b} \Omega_{\rr b} \delta_{\rr b} +\beta_{\rr c} \Omega_{\rr c} \delta_{\rr c}}{k^2 +m^2 a^2}~.
\ee
The baryon contrast satisfies
\be \label{eq:deltabMG}
\delta_{\rr b}'' +{\cal H} \delta_{\rr b}' +\left[  c_{\rr b}^2 k^2 - \frac{3}{2} {\cal H}^2 \Omega_{\rr b} \left(1+\frac{2\beta_{\rr b}^2}{1+ \frac{m^2a^2}{k^2}}\right) \right] \delta_{\rr b}-\frac{3}{2} {\cal H}^2\Omega_{\rr c} \left(1+\frac{2\beta_{\rr b}\beta_{\rr c}}{1+ \frac{m^2a^2}{k^2}}\right)\delta_{\rr c}=0~,
\ee
and CDM
\be \label{eq:deltaDMMG}
\delta_{\rr c}'' +{\cal H} \delta_{\rr c}' - \frac{3}{2} {\cal H}^2 \Omega_{\rr b} \left(1+\frac{2\beta_{\rr b}\beta_{\rr c}}{1+ \frac{m^2a^2}{k^2}}\right)\delta_{\rr b}-\frac{3}{2} {\cal H}^2\Omega_{\rr c} \left(1+\frac{2\beta_{\rr c}^2}{1+ \frac{m^2a^2}{k^2}}\right)\delta_{\rr c}=0~.
\ee
These modified equations allow one to study the influence of modified gravity on 21 cm physics.   The relative differences between the baryon perturbation evolution for the $\Lambda$-CDM model and for some models of modified gravity have been plotted in Fig.~\ref{fig:baryonevol} for the wavelength mode $k=0.1 \Mpc^{-1} $ in the observable range of 21cm experiments.

\subsection{Some Models}

\subsubsection{ Chameleons}

We will consider chameleon models where the coupling to matter $\beta$ is constant and the potential $V(\phi)$ is a decreasing function of $\phi$. As a result the effective potential
\be
V_{\rm eff}(\phi)= V(\phi) +e^{\beta_{\rr b} \phi/m_{\rm Pl}}\rho_{\rr b} + e^{\beta_{\rr c} \phi/m_{\rm Pl}} \rho_{\rr c}~,
\ee
where the couplings to baryons  and CDM are taken to differ, has a minimum $\phi_{\rm min}$. When the mass of the scalar field at this minimum $m$ is large enough and
greater than the Hubble rate $H$, the minimum is stable since before Big Bang Nucleosynthesis (BBN), guaranteeing a small variation of the masses of particles in the Einstein frame, and therefore a negligible modification of the formation of matter during BBN.
This condition must be satisfied also to avoid important deviations from  $\Lambda$-CDM at the homogeneous level.

Typical examples of potentials correspond to inverse power laws of the Ratra-Peebles types.
Local experiments such as cavity tests of gravity or the Lunar Ranging test of the equivalence principle impose stringent restrictions on the mass of the scalar field now. We will come back to these restrictions later.

\subsubsection{f(R) models}

Viable $f(R)$ models are nothing but chameleon models with a definite value of the coupling $\beta=1/\sqrt{6}$, and with the potential given by
\be
V(\phi)= m_{\rm Pl}^2 \frac{Rf_R-f(R)}{2f_R^2}~,
\end{equation}
where $f_R=\dd f / \dd R$. The mapping between $R$ and $\phi$ is given by
\be
f_R=e^{-2\beta \phi/m_{\rm Pl}}~.
\ee
Typically we will be interested in large curvature models $R\gtrsim R_\star$, where
\be
f(R)= R + R_0 +R_1 \left(\frac{R_\star}{R}\right)^n
\ee
and $R_0$ plays the role of a cosmological constant and $n>0$ defines the asymptotic expansion of the $f(R)$ function. For these models, the range of the scalar force
is constrained by the requirement $m_0/H_0 \gtrsim 10^3$ which springs from the loose constraint that galaxies such as the Milky Way should have a thin shell.

\subsubsection{Symmetrons}

For the symmetron model, the interaction potential and the
coupling function are simply chosen such that
\begin{eqnarray}
V(\phi) &=& V_{0}-\frac{1}{2}\mu_{\rr {sym}}^2 \phi^2 + \frac{1}{4}\lambda_{\rr {sym}}
\phi^4~,\nonumber\\
A(\phi) &=& 1 + \frac{\phi^2}{2M^2}~,
\end{eqnarray}
in which $V_{0}$ is a cosmological constant.
The
effective potential can be rewritten as
\begin{eqnarray}\label{pot}
V_{\rm eff}(\phi) &=& \frac{1}{2}\left(\frac{\rho_{\rm
m}}{M^2}-\mu_{\rr{sym}}^2\right)\phi^2+\frac{1}{4}\lambda_{\rr {sym}} \phi^4~.
\end{eqnarray}
Hence, in the symmetron model, as long as $\rho_{\rm m}$ is high
enough, namely $\rho_{\rm m}\ge\rho_{\star}$ where
\begin{eqnarray}
\rho_\star &\equiv& \mu_{\rr {sym}} ^2 M^2~,
\end{eqnarray}
the minimum of the effective potential is at the origin
($\phi=0$). In contrast, in vacuum, the symmetry is broken and the
potential has two nonzero minima:
\begin{eqnarray}
\phi_\star &=& \pm\frac{\mu_{\rr {sym}} }{\sqrt{ \lambda_{\rr {sym}}} } ~.
\end{eqnarray}
When the matter density does not vanish, the minimum of the effective potential depends on $\rho_{\rr m}$:
\begin{eqnarray}
\phi_{\rm min}(\rho_{\rr m}) = \phi_{\star} \sqrt{1 - \frac{\rho_{\rr m}}{\rho_{\star}}} \theta\left(\rho_{\star} -  \rho_{\rr m}\right)~,
\end{eqnarray}
where $\theta(x)$ is the Heaviside function. The minimum is not zero only for $\rho_{\rr m}<\rho_*$ and converges to $\phi_\star$ when $\rho_{\rr m}$ vanishes.
As long as $\phi \ll M$ the effective coupling to matter reads
\begin{eqnarray}
\beta(\phi)\
\approx\ \frac{m_{\rm{Pl}}\phi}{M^2}~,
\end{eqnarray}
leading to the absence of modification of gravity in dense
environments where the field vanishes. Indeed, the variation of the effective coupling with the matter density is:
\begin{equation}
\beta_{\rm min}(\rho_{\rr m}) = \beta_{\star} \left(1-\frac{\rho_{\rr m}}{\rho_{\star}}\right) \theta\left(\rho_{\star} - \rho_{\rr m}\right)~.
\end{equation}

The symmetron model is designed to induce modifications of gravity
which could be tested in the near future, both gravitationally and
cosmologically. Requiring that the energy density at which the
curvature at the origin of the potential changes sign
is close to the current critical energy density , we have the
estimate
\begin{eqnarray}
M^2 \mu_{\rr{sym}}^2 &\sim& H_0^2 m_{\rr{Pl}}^2~.
\end{eqnarray}
Moreover, the modification to gravity is detectable only if it is
comparable to (or bigger than)  standard gravity, or
equivalently the effective coupling
$\beta\sim\mathcal{O}(1)$, which implies
\begin{eqnarray}
\frac{\phi_\star}{M} &\sim& \frac{1}{\sqrt{ A_2}}~,\nonumber
\end{eqnarray}
where we have defined $A_2\equiv m_{\rm Pl}^2/M^2$. These
determine the vacuum mass
\begin{eqnarray}
m^2(\phi_0)\ =\ 2\mu_{\rr{sym}}^2\ \sim\
\mathcal{O}\left(\frac{m_{\rr{Pl}}^2}{M^2}\right)H_0^2\ =\
\mathcal{O}(A_2)H^2_0
\end{eqnarray}
and correspondingly set the interaction range of the symmetron to
be $\sim\mathcal{O}\left[m^{-1}(\phi_0)\right]$.
From the study of solar system tests, it can been found that $M\lesssim 10^{-3}m_{\rr{Pl}}$ or equivalently
\begin{equation}
A_2 \gtrsim 10^6.
\end{equation}
This implies that $\phi_\star / M \lesssim 10^{-3}$.
The range of the symmetron in the cosmological  vacuum is given by
\begin{eqnarray}
\mu_{\rr{sym}}^{-1}\lesssim 10^3H_0^{-1}\sim 10\ \rr{Mpc}~,
\end{eqnarray}
which corresponds to relevant scales for astrophysics.

\subsubsection{Dilatons}

Dilaton models are akin to symmetrons in as much as the effective potential has a minimum whose origin, here, springs from the coupling to matter
\be
A(\phi)= 1+ \frac{A_2}{2 m_{\rm Pl}^2} (\phi-\phi_\star)^2~,
\ee
which leads to a universal coupling
\be
\beta (\phi)= A_2( \phi-\phi_\star)~.
\ee
On the contrary, the bare potential $V(\phi)$ is assumed to be a smooth and non-vanishing function  whose order of magnitude
$V(\phi_\star)$ corresponds to the vacuum energy now. The dynamics can be analysed close to $\phi=\phi_\star$ where the field is attracted in high density environments. Local constraints on modifications of gravity impose that
\be
A_2 \gtrsim 10^6~,
\ee
in a similar fashion to the symmetron case, which implies here that $m_0/H_0 \gtrsim 10^3$ again.

\subsection{Reconstructing the Dynamics} \label{sec:reconstruction}
All the models which have been presented in the previous section can be parameterised solely using the time evolution of the mass and coupling functions at the minimum
of the effective potential. This is a particularly convenient way of defining models when one is interested in cosmological perturbations.
The dynamics of the models of modified gravity of the chameleon, dilaton and symmetron types are completely determined by the minimum equation of the effective potential
\be
\left. \frac{\dd V}{\dd \phi}\right|_{\phi_{\rr{min}}}= -\beta \frac{\rho}{m_{\rm Pl}}~,
\end{equation}
where we have introduced $\rho= 3 H^2 m_{\rm Pl}^2$,
\be
\beta=\beta_{ \rr c} A_{ \rr c} \Omega_{\rr c} + \beta_{\rr b} A_{\rr b} \Omega_{\rr b}
\ee
and $A_{\rr{b,c}}\sim 1$ due to BBN constraints.
In fact, the knowledge of the time evolution of the mass $m$ and the coupling $\beta$ is enough to determine the time evolution of the field and the potential completely.
Using the minimum equation
we deduce that the field evolves according to
\be
\frac{\dd \phi}{\dd t}=\frac{3H}{m^2} \beta  \frac{\rho}{m_{\rm Pl}}.
\ee
This is the time evolution of the scalar field at the background level since the instant when the field starts being at the minimum of the effective potential.
This leads to the solution
\begin{equation}
\phi(a)=  \frac{3}{m_{\rm Pl}}\int_{a_{\rm ini}}^a \frac{\beta (a)}{a m^2(a)}\rho (a)  \dd a +\phi_{\rr c}~,\label{phi}
\end{equation}
where $\phi_{\rr c}$ is the initial value of the scalar field.
Similarly the minimum equation implies that the potential can be reconstructed as a function of time
\begin{equation}
V=V_0 -3  \int_{a_{\rm ini}}^a \frac{\beta(a)^2}{am^2(a)} \frac{\rho^2}{m^2_{\rm Pl}} \dd a,
\label{V}
\end{equation}
where $V_0$ is the initial value of the potential at $a=a_{\rm ini}$.
This defines the bare scalar field potential $V(\phi)$ parametrically when $\beta (a)$ and $m(a)$ are given. Let us come back to the chameleon,  $f(R)$, dilaton and symmetron models for which one can explicitly verify that this way of defining models can be used.

\subsubsection{Chameleon and f(R) models}
For these models and in the matter era, the coupling to matter $\beta$ is constant. We will be interested in the models where
\be
m=m_0 a^{-r}~.
\ee
When
$r>3$ and $\beta=1 /\sqrt 6$, they correspond to the large curvature $f(R)$ models with $r=3(n+2)/2$ in the matter era. When $3/2<r<3$, the models are of the chameleon type with an inverse power law potential
\be V(\phi) \sim \frac{\Lambda^{n+4}}{\phi^n}~,\ee
and $n=(2r-6)/(2r-3)$.

\subsubsection{Symmetron}

The symmetron models can be reconstructed using
\be
\beta(a)=  \beta_\star \sqrt{1-\left(\frac{a_\star}{a}\right)^3}
\ee
for $z<z_\star$ and $\beta=0, \ z>z_\star$. Similarly we have
\begin{equation}
m(a)=m_\star\sqrt{1-\left(\frac{a_\star}{a}\right)^3}~.
\ee
The parameters $(\beta_\star,m_\star,z_\star)$ determine the model completely,
\be
\phi_\star= \frac{2\beta_\star \rho_\star}{m_\star^2 m_{\rm Pl}}, \hspace{2mm}Ê
m_\star= \sqrt {2} \mu_{\rr{sym}}, \hspace{2mm} \lambda_{\rr{sym}}=\frac{\mu_{\rr{sym}}^2}{\phi_\star^2}~.
\ee
Finally we have
\be
\beta(\phi)= \frac{\beta_\star}{\phi_\star} \phi~.
\end{equation}

\subsubsection{Dilatons}
For dilatons, the behaviour of the mass and coupling functions in the matter era close to $\phi_\star$ is universal and defined by
\be
m=m_0 a^{-2}~,
\ee
corresponding to a mass which is proportional to the Hubble rate. The coupling to matter is time dependent and follows the inverse matter density
\be
\beta = \beta_0 a^{3}~,
\ee
with a coupling which increases as matter becomes sparser. This model can be seen as a generalisation of chameleon models where one can choose
\be
m=m_0 a^{-r},\ \beta= \beta_0 a^{-s}~,
\ee
where $r=2$ and $s=-3$ here. We will consider these $(r,s)$ models in the following.

\section{21cm power spectrum of modified gravity}  \label{sec:21cmeffects}

\subsection{21cm FFTT-type experiment}  \label{sec:FFTT}

The present and  next generations of large radio-telescopes, like LOFAR~\cite{Garrett:2009gp, Harker:2010ht}, MWA~\cite{Mitchell:2010cc} and SKA~\cite{Lazio:2009bea,Rawlings:2011dd} are designed for the detection of the 21cm signal from the period of reionisation.   However, their sensitivity is not expected to be sufficient to constrain cosmology\footnote{Except for the SKA and for some optimistic scenarios~\cite{Mao:2008ug}} through the observation of the 21cm power spectrum~\cite{Mao:2008ug}.   Nevertheless, M. Tegmark and M. Zaldarriaga have recently proposed the concept of Fast Fourier Transform radio-Telescopes (FFTT)~\cite{Tegmark:2008au} whose potential ability to measure the 21cm power spectrum and to put strong constraints on the cosmological parameters have been demonstrated in Ref.~\cite{Mao:2008ug}.   The FFTT is an all digital radio-telescope composed of a square grid of dipole antennas.   The multifrequency images of half the sky are reconstructed from the data measured by each antenna after several fast Fourier transforms.  The key advantage of the FFTT compared to traditional interferometric radio-telescopes is its cost, scaling as $\Nant \log_2 \Nant$ instead of $ \Nant^2$, with $\Nant$ the number of dipole antenna.

 In this paper, we adopt the FFTT as a template to study modified gravity effects on the 21cm power spectrum from  reionisation.
We refer to Refs.\cite{Tegmark:2008au,Mao:2008ug} for the main specifications of the experiment that are reported in Tab.~\ref{tab:FFTT}.  The aim of this section is to evaluate the errors on the 21cm power spectrum for the considered FFTT experiment.  For simplicity, we assume ideal foreground removals and refer the interested reader   in foregrounds and removal techniques  to Refs.~\cite{Liu:2011hh,Liu:2011ih,Petrovic:2010me}.

The noise spectrum $P_{\rr n} $ for the FFTT  is given by~\cite{Tegmark:2008au}
\begin{equation}
P_{\rr n} (\mathbf{ u}) = \frac{4 \pi  f_{\rr{sky}}\lambda^2 T_{\rr{sys}} ^2
}{f_{\rr{cover}}^2 D_{\rr{max}}^2
  \Omega_{\rr{fov}} t_{\rr o} } P_{\rr n}^{\bot} (u_\bot) P_{\rr n
}^{\parallel} (u_\parallel)~,
\label{eq:Pn}
\end{equation}
where $  T_{\rr{sys}} $ is the system temperature, $D_{\rr{max}}$ is
the length of a side of the FFTT, $\Omega_{\rr{fov}}= 2 \pi $ is the total field of view,
$f_{\rr{sky}}=  \Omega_{\rr{fov}}/4\pi$ and $t_{\rr o}$ is the total
observation time. $f_{\rr{cover}}=\Nant (D/ \Dmax)^2$ is the fraction of the total area covered by
the antenna.  $D$ is the minimum baseline between two antennas.    $P_{\rr n}^{\bot}$ and $P_{\rr n}^{\|}$ are the fourier transform in the $\mathbf u$-space of the gaussian angular and frequency window functions.  These are introduced to take into account respectively the angular resolution and the frequency resolution of the telescope.  They read
\begin{eqnarray}
P_{\rr n}^\bot  (u_\bot)  & = & \rr e^{ \mathbf u_\bot ^2 \sigma_\bot^2}~,\\
P_{\rr n }^\parallel (u_\parallel) & = & \rr e^{ u_\parallel ^2 \sigma_\parallel ^2}~,
\end{eqnarray}
with
\begin{equation}
\sigma_\bot  =  \frac{0.89 \lambda }{ D_{\rr {max}} \sqrt{8 \ln 2}}, \hspace{3mm}
\sigma_\parallel  =  \Delta \nu ~,
\end{equation}
and where  $ \Delta \nu $ is the frequency bandwidth.
Moreover, to avoid non-linear effects, we follow Ref.~\cite{Mao:2008ug} and assume a sharp cut off at $k_{\rr{max}} = 2 \Mpc^{-1} $, so that for our specifications of the FFTT, we can safely consider $P_{\rr n }^\parallel = 1$.

It turns out that the error on the power spectrum for a given mode $\bf u$ reads
%the square root of the inverse Fisher matrix of the experiment,
\be
\delta P_{\Delta \Tbr} ({\bf{u}}) =    \frac 1 {\sqrt{2 N_{\rr c}} }\left[ P_{ \Delta \Tbr} ({\bf {u}} ) + P_{\rr n}  (u_\bot ) \right]~,
\ee
where
\be
N_{\rr c} = \frac{\Omega_{\rr{fov}}}{\Omega_{\rr{patch}}} \frac{2 \pi u_\bot}{\Delta u_\bot}
\ee
 is the number of cells in  $\mathbf u$-space probed by the experiment.   For the flat sky approximation to be valid, it is required to divide the sky in small patches whose size $\Omega_{\rr{patch}}$ is typically smaller than $1$sr, so that $\Delta u_\bot = 2 \pi / \sqrt{\Omega_{\rr{patch}}} $.

 Because the FFTT is a signal-dominated experiment in the range of wavelengths of interest for cosmology, the error on the power spectrum is mainly due to the cosmic variance.  It is noticeable that compared to CMB experiments which only probe the last scattering surface, the induced errors on the cosmological model parameters can in principle be reduced by the observations of the 21cm 3D power spectrum at several redshifts.

\vspace{3mm}
\begin{table}  \begin{center}  \label{tab:FFTT}
\begin{tabular}{|c|c|}
\hline
Total size & $\Dmax=1$ km  \\
Min. baseline & D= 1 m \\
Number of antennas & $\Nant = 10^6$  \\
Bandwidth & $\Delta \nu = 0.05$ Mhz \\
System temperature & $T_{\rr{sys} } = 400 \ \rr K  $ \\
Observation time & $t_{\rr o} = 1$ year  \\
Angular resolution & $\theta_{\rr{res}} =  \lambda / D $ (Beam FWHM) \\
Field of view &$ \Omega_{\rr{fov}} = 2 \pi $ \\
\hline
\end{tabular}
\caption{Specifications of the considered FFTT experiment.  $\lambda$ is the redshifted wavelength of the 21cm signal.
We assume a gaussian beam and use the FWHM convention~\cite{Tegmark:2008au}  for the beam width, as well as ideal foreground removal. The FFTT covers half of the sky sphere.  }
\label{tab:FFTT}
\end{center}
\end{table}

\subsection{21cm Spectra and Explicit Models}

In this section, we calculate the effects of modified gravity on the 21cm power spectrum at reionisation and on the matter power spectrum today.  For each model, we give rough bounds on the model parameters to have observable  signatures with the FFTT.  Then these bounds are compared to those obtained with the matter power spectrum and with local tests of gravity.  At the end of the section, we show that in principle the coupling of the scalar field to photons can also be constrained with 21cm observations through the variation of the fine structure constant $\alpha$.

\subsubsection{$f(R)$ models}

\begin{figure}[h!]
\begin{center}
\includegraphics[height=8cm]{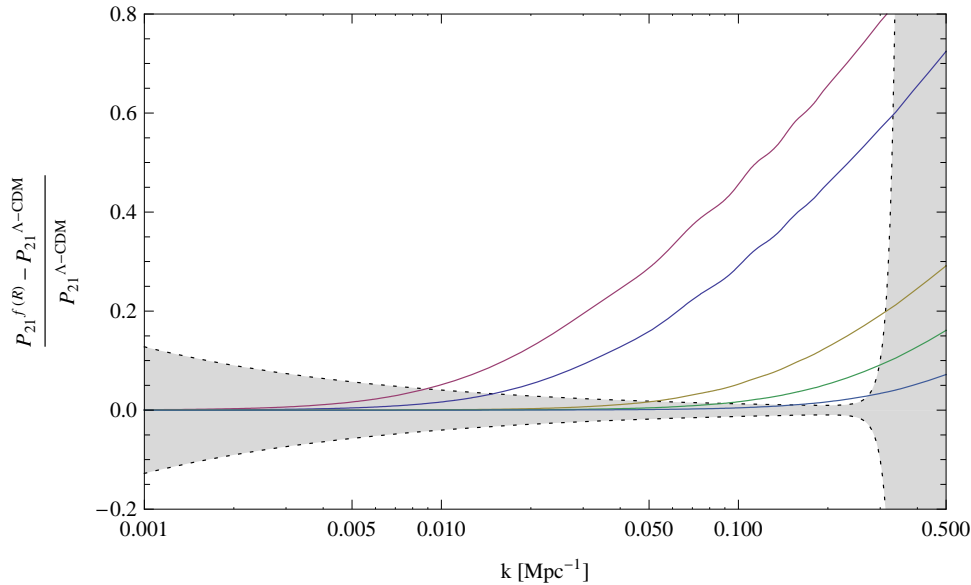}
\caption{Relative differences between the 21cm power spectra for $\Lambda$-CDM and $f(R)$ models, with $\mu = 0$ (i.e. for modes orthogonal to the line of sight).  Power spectra are calculated at $z=11$, assuming a neutral fraction $x_{\rr H} = 0.9$, accordingly to Ref.~\cite{Mao:2008ug}. From top to bottom curves, the model parameter $m_0$ is respectively $5 \times 10^{-5} \Mpc^{-1} $ (red), $10^{-4} \Mpc^{-1} $ (dark blue), $5 \times 10^{-4} \Mpc^{-1} $ (yellow), $ 10^{-3} \Mpc^{-1} $ (green) and $ 2 \times 10^{-3} \Mpc^{-1} $ (blue).  The grey band corresponds to the expected errors on the power spectrum measurements for the considered FFTT experiment.  Errors are due to the cosmic variance at large scales and grow exponentially at small scales due to the angular resolution of the telescope.}
\label{fig:fR}
\end{center}
\end{figure}

The relative differences between the 21cm power spectra at $z=11$ for the $f(R)$ model and for the $\Lambda$-CDM model are plotted in Fig.~\ref{fig:fR} for wavelength modes orthogonal to the line of sight ($\mu = 0$), for $r = 3$, $\beta = 1/\sqrt 6$, and various values of the parameter $m_0$.   The case r=3 corresponds to the limit case $n\to 0$ of large curvature $f(R)$ models.  The expected relative errors on the power spectrum measurements for the FFTT experiment are also added.

On large scales ($k \lesssim 0.01 \Mpc^{-1}$), the terms in  $\beta^2 / (1+ m^2 a^2 / k^2) $ in Eqs.~(\ref{eq:deltabMG}) and (\ref{eq:deltaDMMG}) tend to zero and the $f(R)$ model behaves like the $\Lambda$-CDM model.  The relative difference with the $\Lambda$-CDM model grows at small scales, where these terms become important.  For sufficiently small values of $m_0$ ($\lesssim 10^{-4} \Mpc^{-1}$), the 21cm power spectrum can be strongly affected by modified gravity effects.   For these very low  masses, the mass at $z=11$ is still larger than the Hubble rate. This is not the case now implying that the scalar field does not follow the minimum of the effective potential in the recent past of the Universe, leading to potential deviations from $\Lambda$-CDM at the background level at small redshift. These cases are already excluded by local constraints and are only presented for illustration.   A difference with the $\Lambda$-CDM could be detected by a FFTT experiment up to $m_0 \simeq 2 \times 10^{-3} \Mpc^{-1}$ for modes orthogonal to the line of sight.

\begin{figure}[h!]
\begin{center}
\includegraphics[height=8cm]{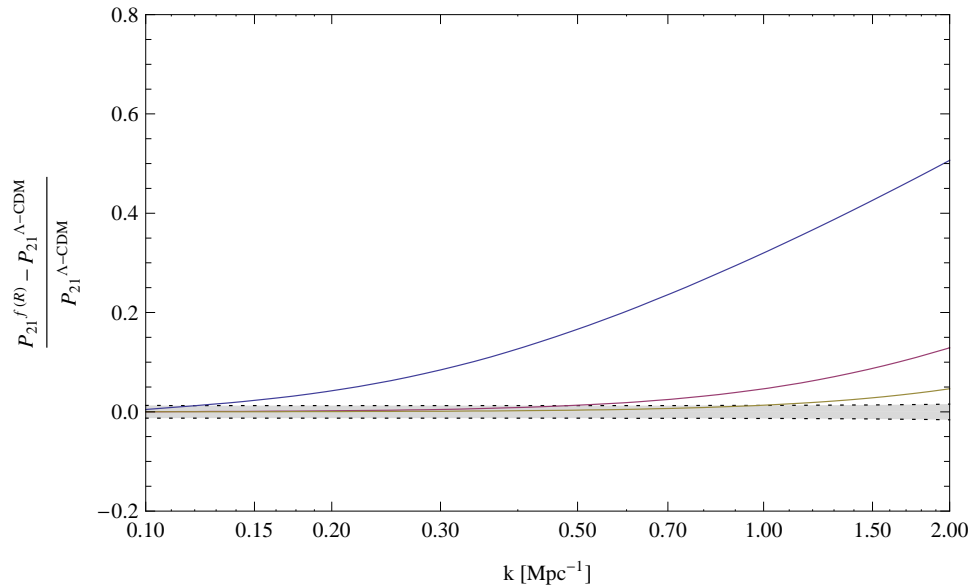}
\caption{Relative differences between the 21cm power spectra for $\Lambda$-CDM and $f(R)$ models, with $k_\bot = 0.1 \Mpc^{-1}$ and $k_\parallel $ varying up to $ 2 \Mpc^{-1}$ (cut-off introduced to avoid non linear effecs).  As in Fig.~\ref{fig:fR}, power spectra are calculated at $z=11$, assuming a neutral fraction $x_{\rr H} = 0.9$.  From top to bottom, $m_0$ values are respectively $ 2 \times 10^{-3} \Mpc^{-1} $ (blue), $10^{-2} \Mpc^{-1}$ (red) and $2 \times 10^{-2} \Mpc^{-1}$ (yellow).  The grey band corresponds to the expected errors on the power spectrum measurements for the considered FFTT experiment.}
\label{fig:fR2}
\end{center}
\end{figure}

The situation is improved if we consider $k_\bot \approx 0.1 \Mpc^{-1}$ (in order to have a good compromise between a large value of $N_c $ and $P_{\rr n}^\bot \simeq 1$), and $k_\parallel $ up to its maximal value $k_{\rr{max}} = 2 \Mpc^{-1}$ (remember that the noise is independent of $k_\parallel$).  The last bound can be enhanced by one order of magnitude in this case, up to $m_0 \simeq 2 \times 10^{-2} \Mpc^{-1}$, as shown in Fig.~\ref{fig:fR2}.  Hence, by using
the frequency component of the 3D-21cm power spectrum, one probes smaller perturbation scales for which the effects of modified gravity are potentially more important.   Such modes cannot be probed directly with the matter power spectrum today since they correspond to non-linear scales for which the screening mechanism suppresses the deviations from the $\Lambda$-CDM behaviour. 

This limit on the scalar field mass corresponds to a ratio $m_0 / H_0 \simeq 90$, \textit{i.e.} lower than the constraint obtained with local tests~\cite{Brax:2011aw}, $m_0 / H_0 \gtrsim 10^3 $ (imposed to guarantee that galaxies such as the Milky Way have a thin shell).  However, it is comparable to the solar system constraint~\cite{Brax:2011aw},
\be \label{eq:solartests}
\frac{m_0^2}{H_0^2} \gtrsim \frac{\beta_0^2 \Omega_{\rr m 0} 10^{-4r + 4s + 12} }{(2 r -s-3) \Phi_{\odot}}~,
\ee
where $\Phi ~\sim 10^{-6}$ is the solar Newtonian potential, giving rise for $s=0$, $r=3$, $\beta_0 = 1/\sqrt 6 $ to the bound $m_0 / H_0 \gtrsim 100$.

We have also calculated the matter power spectrum today for $m_0  = 2 \times 10^{-3} \Mpc^{-1}$  in the linear approximation.  It is plotted in Fig.~\ref{fig:Pk}.   The matter power spectrum is observed to be slightly outside the 68\% C.L. error bars of the SDSS data.  For increasing values of the scalar field mass, the difference with the $\Lambda$-CDM matter power spectrum becomes undetectable.    However, it must be noticed that we did not take into account potentially important non-linear effects.

To summarise, we have found that future observations of the 21cm signal at reionisation with a FFTT-like radiotelescope should constrain more efficiently  the $f(R)$ model with $r=3$ than the present matter power spectrum, mainly because the signal can probe smaller scales in the linear regime.  Modified gravity effects could be detectable up to a scalar field mass value $m_0 \simeq 2 \times 10^{-2} \Mpc^{-1}$, corresponding to $m_0 / H_0 \simeq 90$.  Although this value is lower than the galaxy constraint from \cite{Brax:2011aw}, it is competitive with the tests of gravity in the solar system and better than the bounds from the CMB and other local tests of gravity.  It will therefore be interesting to study whether this situation could be improved by using multi-redshifts measurements and by combining data from the whole 3D $\bf u$-space, or made worse due to possible degeneracies with cosmological and nuisance reionisation parameters.   Such a study will require ideally the use of more complex Fisher matrix or Monte-Carlo methods and is left for future work.   This last remark is also valid for the other models we consider in this paper.

\subsubsection{Symmetron}

\begin{figure}[h!]
\begin{center}
\includegraphics[height=8cm]{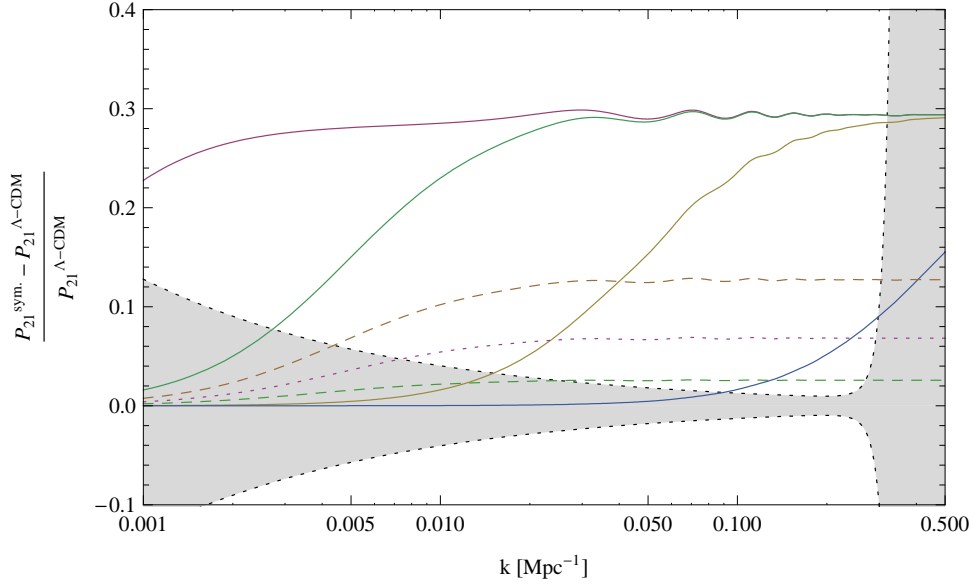}
\caption{Relative differences between the 21cm power spectra for $\Lambda$-CDM and symmetron models.  As in Fig.~\ref{fig:fR}, the power spectra are calculated at $z=11$, assuming $x_{\rr H} = 0.9$, for wavelength modes orthogonal to the line of sight.   Plain curves are for $z_\star = 20$, $\beta_0 = 1$, with $m_0$ varying.  From top left to bottom right, $m_0$ takes the values $10^{-2} \Mpc^{-1}$ (red), $0.1 \Mpc^{-1}$ (green), $1 \Mpc^{-1}$ (yellow) and $10 \Mpc^{-1}$ (blue).
The two dashed curves are for $\beta_0 = 1$, $m_0= 0.1 \Mpc^{-1}$ with $z_\star$ varying.  The top yellow one is for $z_\star = 17$, the bottom green one is for $z_\star = 14$.  The dotted curve is for $z_\star = 20$, $m_0 = 0.1 \Mpc^{-1}$, $\beta_0 = 0.5$.   The parameters $\beta_0$ and $z_\star$ are observed to control the magnitude of the shift between symmetron and $\Lambda$-CDM at small scales, whereas the parameter $m_0$ controls the scale of the transition regime.   The grey band corresponds to the expected errors on the power spectrum measurements for the considered FFTT experiment. }
\label{fig:symmetron}
\end{center}
\end{figure}

\begin{figure}[h!]
\begin{center}
\includegraphics[height=8cm]{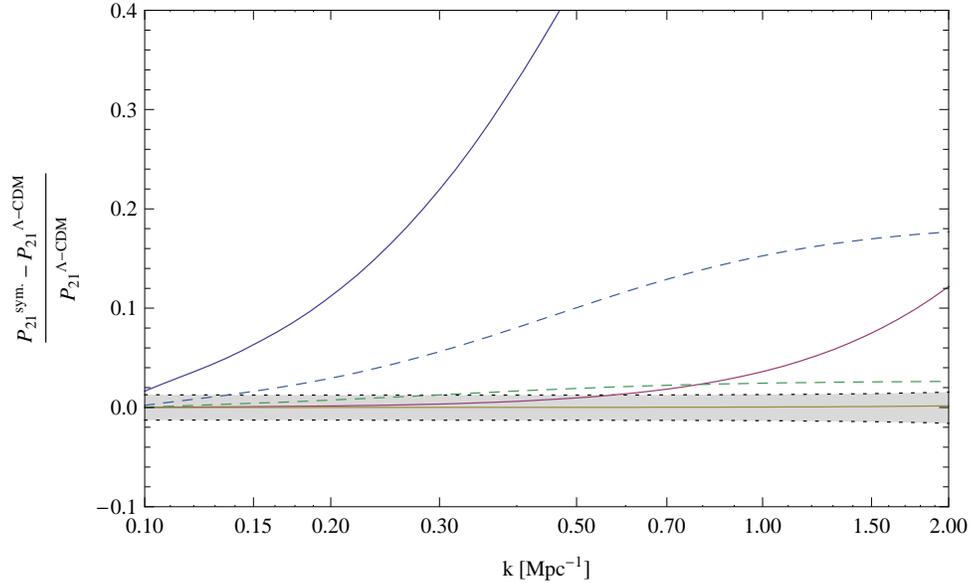}
\caption{Relative differences between the 21cm power spectra for $\Lambda$-CDM and symmetron models, at $z=11$, assuming $x_{\rr H} = 0.9$, for wavelength modes with $k_\bot = 0.1 ~\Mpc^{-1}$.   Plain curves are for $z_\star = 20$, $\beta_0 = 1$ and $m_0 =10/100/1000 ~\Mpc^{-1}$ (respectively the blue, red and yellow plain curves).   The two dashed curves are for $\beta_0 = 1$, $m_0= 10 ~\Mpc^{-1}$, $z_\star=14$ (top blue) and $z_\star = 12$ (bottom green).  The grey band corresponds to the expected errors on the power spectrum measurements for the considered FFTT experiment. }
\label{fig:symmetron2}
\end{center}
\end{figure}

The relative differences between the 21cm power spectra at $z=11$ for the symmetron and the $\Lambda$-CDM models are plotted in Figs.~\ref{fig:symmetron} and \ref{fig:symmetron2}, respectively for $k_\parallel = 0$ and $k_\bot $ varying, and for $k_\bot = 0.1 \Mpc^{-1}$ and $k_\parallel$ varying.
% for various values of  $m_0$,  $\beta_0$ and $z_*$.
At large scales, as for the $f(R)$ model, one gets $\beta^2 / (1+ m^2 a^2 / k^2) \rightarrow 0 $ in Eqs.~(\ref{eq:deltabMG}) and (\ref{eq:deltaDMMG}) and the symmetron cannot be distinguished from the $\Lambda$-CDM model.  At small scales, $\beta^2 / (1+ m^2 a^2 / k^2) \rightarrow \beta^2$, inducing a scale invariant shift of the 21cm power spectrum amplitude.

In the case of orthogonal modes to the line of sight, Fig.~\ref{fig:symmetron} shows that the transition between these two regimes occurs in the range of observable scales for $10^{-2} \Mpc^{-1}< m_0 < 10 \Mpc^{-1}$.  The magnitude of the shift is controlled by $\beta_0  $ and by the redshift $z_*$ from which modifications of gravity are triggered.  For $\beta_0 \simeq 1$, modified gravity effects could be detected by the FFTT up to $z_* \simeq 14$, i.e. just before the reionisation.   However, only values of $\beta_0$ of the order of unity or higher will be detectable.

When parallel modes are probed with $k_\bot \approx 0.1 \Mpc^{-1}$ and $k_\parallel < 2 \Mpc^{-1}$, the transition regime is detectable up to  $m_0 \approx 200 \Mpc^{-1}$ (see Fig.~\ref{fig:symmetron2}).   Moreover, the amplitute of the shift increases since the 21-power spectrum goes like $(1+\mu^2)^2 $ with $\mu \approx 1.1$ at the smallest scales (it is larger than unity due to modified gravity effects in Eq.~\ref{eq:mu}).   As a result, the model signatures could be detectable up to $z_* \approx 12$, i.e. just before the redshit of observation, provided $\beta_0 \approx 1$ or higher.

\begin{figure}[]
\begin{center}
\includegraphics[height=8cm]{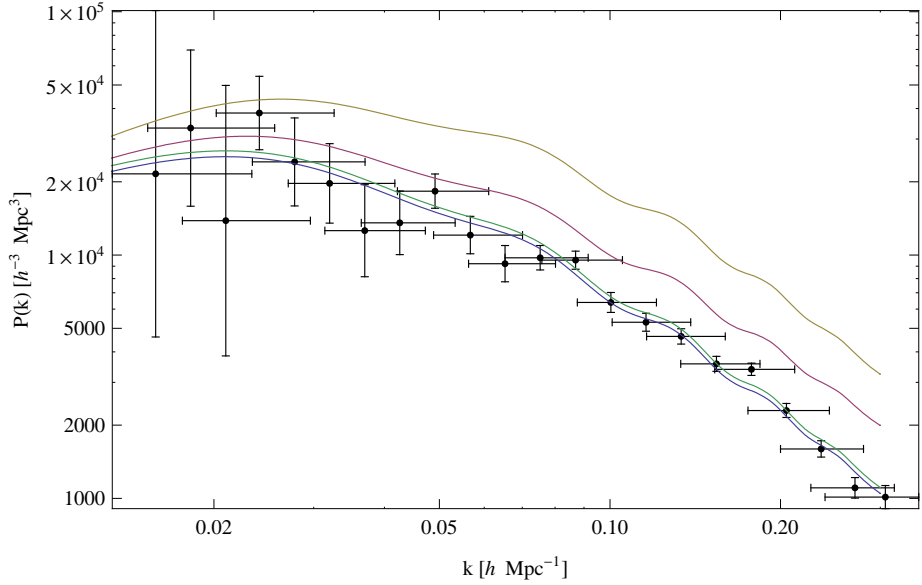}
\caption{Matter power spectrum in the linear approximation for the $\Lambda$-CDM model (blue curve), for the $f(R)$ model with $m_0 = 2 \times 10^{-3} \Mpc^{-1}$ (red curve), for the symmetron model with $z_*= 20$, $\beta_0 = 0.5$ and $m_0 = 0.1 \Mpc^{-1}$ (yellow curve) and for the dilaton model with $m_0 = 10^{-3} \Mpc^{-1}$ and $\beta_0 = 1$ (green curve).  SDSS data points and error bars are also plotted.  }
\label{fig:Pk}
\end{center}
\end{figure}

Signatures on the matter power spectrum today can also be important for the symmetron model, as shown in Fig.~\ref{fig:Pk}  for $\beta = 0.5$, $m_0 = 0.1 \Mpc^{-1}$ and $z_* = 20$.   Typically, values of $m_0 < 10 \Mpc^{-1} $ with $\beta_0 \simeq 1$ and $z_* \gtrsim 10 $ are already ruled out by observations.   Nevertheless, as noticed above for the $f(R)$ model, the matter power spectrum is limited to scales $k \lesssim 0.3 \Mpc^{-1} $ and as a consequence its ability to probe large values of $m_0$ is reduced compared to the 21cm signal.

Local constraints for the symmetron model are satisfied provided $m_0 / H_0 \gtrsim 10^3$~\cite{Brax:2012gr}.   The matter power spectrum in the linear approximation gives a bound on $m_0$ for $\beta_0 \approx 1 $ and $z_\star \approx 20$ that is of the same order of magnitude.  This bound could be improved by observations of the 21cm signal.

To summarise, provided that symmetron effects are triggered at redshifts larger than the redshifts of observation, the 21cm signal is found to be promising to put  stringent constraints on the symmetron parameters, and especially the scalar field mass today,  than the matter power spectrum and the local test of gravity  ($m_0 \lesssim 200\ {\rm Mpc}^{-1}$ for $z_\star = 20 $ and $\beta_0\sim \mathcal O(1)$, i.e. approximatively three order of magnitudes better than local test constraints). Combining those signals and methods could be also a way to break the degeneracy between the model parameters (especially $\beta_0 $ and $z_\star$) by probing different stages of the evolution of the matter perturbations and different environments.

\subsubsection{Dilaton}

For dilaton models, $m(a) = m_0 a^{-2} $ and $\beta(a) = \beta_0 a^3 $ in the matter dominated era, with typically $\beta_0$ of order unity.  One thus gets $\beta^2 / (1+ m^2 a^2 / k^2) = \beta_0 a^{3} / [1+ m_0^2  / (a^2 k^2) ]$ in Eqs.~(\ref{eq:deltabMG}) and (\ref{eq:deltaDMMG}) for the evolution of matter perturbations.   At redshifts $z \simeq 10 $, i.e. during the reionisation, the coupling to matter was therefore about a thousand times lower than  today, so that the model signatures on the 21cm power spectrum are indistinguishable from the $\Lambda$-CDM case if we want the matter power spectrum today to remain under control.   For large values of the scalar field mass (typically $m_0 \gtrsim 0.1 \Mpc^{-3} $), the difference with the $\Lambda$-CDM model is even more suppressed since $m_0^2 / (a^2 k^2) \gg 1$.

\subsubsection{Chameleon models}

\begin{figure}[]
\begin{center}
\includegraphics[height=8cm]{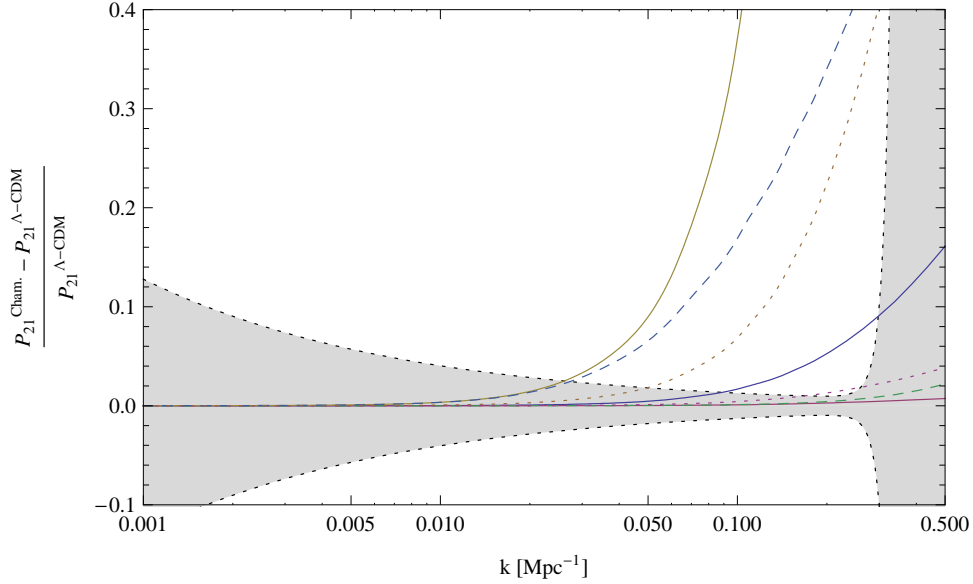}
\caption{Relative differences between the 21cm power spectra for $\Lambda$-CDM and chameleon models, with $k_\parallel = 0$.  As in Fig.~\ref{fig:fR}, the power spectra are calculated at $z=11$, assuming $x_{\rr H} = 0.9$.  Fiducial parameter values are $m_0 = 10^{-3} \Mpc^{-1}$, $\beta_0 = 1/\sqrt 6$, $r= 3$ and $s=0$.   Plain curves are for $s = 0.5$ (top yellow), $s=0$ (blue) and $s= -0.5$ (bottom red).  Dashed curves are for $r=3.5$ (bottom green) and $r=2.5$ (top blue) and dotted curves are for $\beta_0 = 2 / \sqrt{6}$ (top yellow) and $\beta_0 = 1/ (2 \sqrt 6 ) $ (bottom red).
The grey band corresponds to the expected errors on the power spectrum measurements for the considered FFTT experiment. }
\label{fig:chameleon}
\end{center}
\end{figure}

\begin{figure}[]
\begin{center}
\includegraphics[height=8cm]{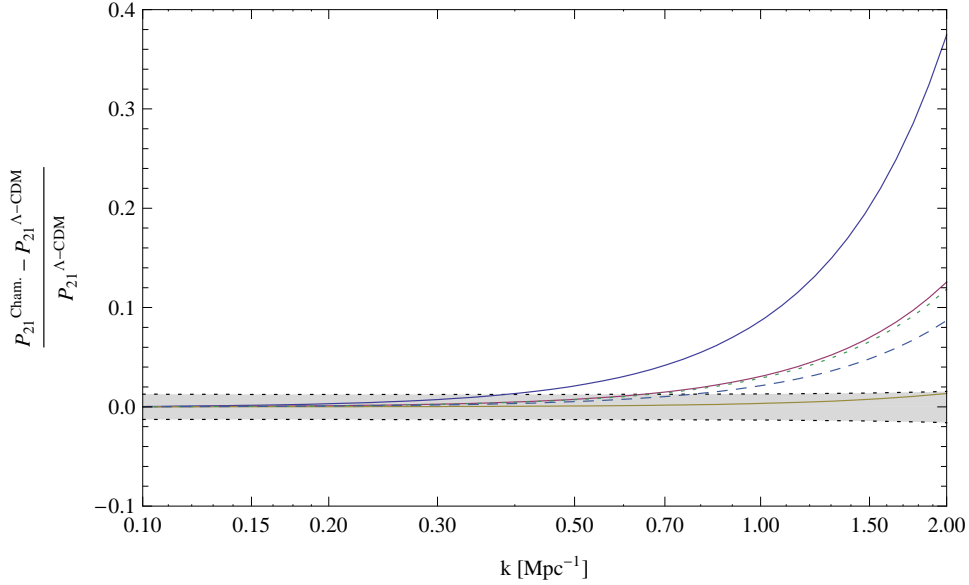}
\caption{Relative differences between the 21cm power spectra at $z=11$, assuming $x_{\rr H}=0.9$, for $\Lambda$-CDM and chameleon models, with $k_\bot = 0.1 \Mpc^{-1}$.
Model parameters for the plain curves are $m_0 = 0.1 \Mpc^{-1}$, $r=3$, $s=0$ and $\beta_0 = 5$ (top blue), $ \beta_0 = 3$ (red) and $\beta_0 = 1$ (bottom yellow).  The dotted curve is for $m_0 = 0.1 \Mpc^{-1}$, $s=0$, $\beta_0 = 5$ and $r= 3.2$~.  The dashed curve is for $ s=0.5 $, $ r= 3$, $\beta_0 = 5$ and $m_0 = 0.8 \Mpc^{-1} $ (from Eq.~\ref{eq:m0limit}).
The grey band corresponds to the expected errors on the power spectrum measurements for the considered FFTT experiment. }
\label{fig:chameleon2}
\end{center}
\end{figure}

The dilaton and $f(R)$ models are particular cases of generalised chameleon models, for which $\beta_0$, $m_0$, $r$ and $s$ can vary.
In Figs.~\ref{fig:chameleon} and \ref{fig:chameleon2}, we have calculated the 21cm power spectrum at reionisation for various values of $\beta_0$, $r$ and $s$.  Increasing $\beta_0$ or $s$ increases the relative difference with the $\Lambda$-CDM model while increasing $r$ implies a reduction of this difference.

In the case $r> 3$, $s<0$ and $\beta_0 \sim \mathcal O (1)$, the best constraints from local tests of gravity come from the galaxies, imposing $m_0 / H_0 > 10^3$, i.e. approximatively $m_0 \gtrsim 0.1 \Mpc^{-1}$.   This bound increases by several orders of magnitude if $r < 3$ due to the stringent constraints from laboratory experiments with cavities.   If $s \gtrsim 1$, solar system tests of gravity can give the best constraints with the condition given by Eq.~(\ref{eq:solartests}).

If we impose $ s= 0$ and $r=3$ together with $m_0 = 0.1 \Mpc^{-1}$ (i.e. about the limiting value imposed by galaxy constraints), we have found that the coupling to matter today needs to be $\beta_0 \gtrsim 20 $ to lead to detectable effects in the 21cm power spectrum with $k_\parallel = 0$.   This limit is lowered to $\beta_0 \approx 2$ for modes with $k_\bot = 0.1 \Mpc^{-1}$ and $k_\parallel$ varying.  For negative values of $s$, the limit on $\beta_0 $ increases.
If we take positive values of $s$ and impose $\beta_0 = 1$ as well as $r=3$ and take
\be \label{eq:m0limit}
m_0 = H_0  \sqrt{\frac{\Omega_{\rr m} 10^{ 4s -4r +18}   }{3-s}}
\ee
 (i.e. the minimal value for the solar system constraints to be respected), the effects on the 21cm power spectrum are never observable.  An identical conclusion can be drawn for the matter power spectrum.
Increasing the value of $r$ while the other parameters remain fixed reduces the relative difference with the $\Lambda$-CDM model.   Taking $r<3$ increases this difference but it is compensated by the fact that larger values of $m_0$ are required to respect the bounds from cavity experiments.

To summarise, it appears that the case $r=3$ and $s=0$ is the configuration for which the 21cm signal could reach the sensitivity of the local tests, and signatures could be detected with a coupling to matter $\beta_0 \gtrsim 2$.   Varying $r$ and $s$  leads to more stringent constraints from local tests so that such configurations should be very difficult to probe efficiently with 21cm experiments if the coupling to matter today is of the order of unity or lower.

\subsection{Probing $\beta_\gamma$ via the variation of $\alpha$}

Due to quantum effects such as the presence of heavy fermions, the scalar field $\phi$ can be coupled to photons
\be
\Sgauge = - \frac 1 {4 g^2} \int \dd^4 x \sqrt {-g} B_F(\phi ) F_{\mu \nu} F^{\mu \nu}~,
\ee
where $g$ is the bare coupling constant and $B_F (\phi) = 1 + \beta_\gamma \kappa_4 \phi + \dots $.   We consider the coefficients $\beta$ and $\beta_\gamma $ as free parameters, even if they can be related depending on the model.
The coupling to the electromagnetic field leads to a time dependence of the fine structure constant $\alpha$~\cite{Brax:2011aw,Brax:2012gr},
\be
\frac{1}{\alpha_0} = \frac{1}{\alpha} B_F(\phi)~,
\ee
so that if we assume that $B_F (\phi) \approx 1$, we get
\be \label{eq:dalphaoveralpha}
\frac{\dot \alpha}{\alpha} \approx - \beta_\gamma \kappa_4 \dot \phi~.
\ee
The 21cm radiation is very sensitive to the variation of the fine structure constant, as noticed in Ref.~\cite{Khatri:2007yv}\footnote{Notice however that the results of Ref.~\cite{Khatri:2007yv} has been disputed and have lead to a dispute on how to calculate consistently the effects of the time variation of $\alpha$ on the 21cm signal~\cite{Flambaum:2010um,Khatri:2010ns}.}.   Indeed, the brightness temperature depends on the Einstein coefficient $A_{10}$, itself depending on $\alpha$.  More precisely, $A_{10} =  2 \pi \alpha \nu_{\rr{21}}^3 h_{\rr p}^2 / (3 c^4 m_{\rr e} ^2 )$, and $\nu_{\rr{21} } \propto \alpha^2 R_\infty \propto \alpha^4$, where $R_\infty$ is the Rydberg constant.  Therefore the 21cm brightness temperature $T_{\rr B} \propto A_{10}/ \nu_{21}^2 \propto \alpha^{5}$, and the 21cm power spectrum $P_{\Delta T_{\rr B}} \propto \alpha^{10}$.   As a consequence, the relative error on $\alpha$ from a given experiment will be ten times better than the relative error on the 21cm power spectrum,
\be
\frac {\Delta \alpha}{\alpha} = \frac{1}{10} \frac{\Delta P_{\Delta T_{\rr B}} }{P_{\Delta T_{\rr B}} }~.
\ee
From this argument, one expects that measuring the 21cm signal could be an excellent way to probe the variation of the fine-structure constant, and by extension a probe of  the coupling $\beta_\gamma$.
However, in the context of modified gravity and a signal from the reionisation, the situation is not so ideal.  The coupling to photons not only affects the 21cm brightness temperature through the time variation of the fine structure constant but also through the time-dependent fermion masses.  Moreover, it is not clear how variations of $\alpha$ and fermion masses can affect the reionisation process itself, and thus the time evolution of $x_{\rr H}$, that is fairly unknow even in absence of modified gravity effects.

From these considerations, it appears nearly impossible to constrain the value of $\beta_\gamma$ from single redshift observations only.   Nevertheless, in the context of 21cm tomography over a broad range of redshifts  (typically from the early stages of the reionisation to the period following its completion, $12 \gtrsim z \gtrsim 2$), the evolution of the mean ionized fraction could be reconstructed and combining high and low redshift measurements could provide a natural way of constraining the variation of $\alpha$.   Below, we estimate a bound on $\beta_\gamma$ that could be established from 21cm observations with the FFTT, under the assumption that the mass variation of fermions can be neglected in the brightness temperature.

Assuming that the 21cm power spectrum amplitude will be measured with a percent level accuracy, one can constrain $ |\Delta \alpha| / \alpha \lesssim 10^{-3} $ and a bound on the parameter $\beta_\gamma$ can be derived.  From Eq.~(\ref{eq:dalphaoveralpha}), one gets
\be
 | \beta_\gamma  | \lesssim 10^{-3} \times \frac{1}{\kappa_4 \Delta \phi}~,
\ee
where $\Delta \phi$ is the scalar field variation during the redshift range of observation of the 21cm signal.  It can be calculated from Eq.~(\ref{phi}).
For the $f(R)$, dilaton and chameleon model,  one gets
\be
 \kappa_4 \Delta \phi = \frac{9 \beta_0 H_0^2 \Omega_{\rr m}  }{m_0^2 (2r -s - 2)} \left( a_{\rr{max}} ^{2 r - s- 2}- a_{\rr{min}}^{2 r - s - 2}   \right)~,
\ee
where $a_{\rr{max}} $ and $a_{\rr{min}}$ are respectively the maximal and minimal value of the scale factor in the range of the 21cm observations.  For instance, for the redshift range given above and the $f(R)$ model, one can obtain
\be
 | \beta_\gamma  | \lesssim 0.3 \frac{m_0^2 }{H_0^2}~.
\ee
With $m_0 / H_0 \approx 10^3$, one gets $\beta_\gamma \lesssim 10^6$, which is an intermediate value between the present observational bound $\beta_{\gamma 0} \lesssim 10^{11}$ and the much tighter bound $\beta_{\gamma 0} \lesssim 0.1$ derived in~\cite{Brax:2012gr} from the best experimental bound on the variation of $\alpha$.   The bound on $\beta_\gamma$ remains at a similar order of magnitude for dilaton and chameleon models, for reasonable values of $r$ and $s$ and as long as $\beta_0 = \mathcal O(1)$.

On the other hand, for the symmetron model,
\be
\kappa_4\Delta \phi  = \frac{27 \beta_0 H_0^2 \Omega_{\rr m} }{2 a_*^3 m_0^2 }
\left[ \sqrt{1 - \left(   \frac{a_*}{ a_{\rr{max}}}\right)^3} - \sqrt{1 - \left(   \frac{a_*}{ a_{\rr{min}}}\right)^3}  \right]~.
\ee
As an example, for $z_* = 20$, one gets
\be
\kappa_4 \Delta \phi  \simeq 5 \times 10^{3} \frac{\beta_0 H_0^2}{ m_0^2}~,
\ee
and thus
\be
 | \beta_\gamma  | \lesssim 2 \times 10^{-7}  \frac{ m_0^2 }{\beta_0 H_0^2}~.
\ee
With $m_0 / H_0 \approx 10^3$ and $\beta_0 \approx 1$, one gets $\beta_\gamma \lesssim 0.2$, which is much tighter than the bound obtained for $f(R)$, dilaton and chameleon models.

\section{Conclusion}

The 21 cm line can in principle be used to probe the evolution of the matter perturbations over a wide range of redshifts, typically from the dark ages up to the completion of the reionisation.  Observing the 21cm cosmological signal should therefore further our understanding of the evolution of
the Universe and it is thus important to investigate the predictions of different
cosmological models on the 21cm three-dimentional power spectrum.  In this paper we have considered
modified gravity models with a screening mechanism and study their signatures on the 21cm power spectrum at reionization.

Our archetypical experiment is the Fast Fourier Transform radio-Telescope (FFTT), consisting of a one kilometer side square of dipole antennas, that is designed especially for the detection of the 21cm power spectrum at reionisation.   While the current and next generation of giant radio-telescope are expected to have a limited interest for cosmology, the ability of a FFTT-type experiment to put strong constraints on the various cosmological parameters has been demonstrated in~\cite{Mao:2008ug}.

We have investigated modified gravity models with a screening mechanism
using a unified parametrisation whereby the models are characterised by the
scale dependence of the coupling to matter  and the mass of the scalar field. This has
previously been shown to encapsulate all the effects of modified gravity
with a screening mechanism~\cite{Brax:2011aw} and has been used to
investigate the effects of such models on large scale structure \cite{Brax:2012gr}.

For a fixed redshift, the consequences  of modified gravity are important
on small scales, within the Compton wavelength of the scalar field mediating the deviation from General Relativity.  At the time of reionisation, such small scales could be in the linear regime of perturbations, where modifications of gravity enhance the growth of structure in a maximal way, whereas they are  in the non-linear regime at the higher redshifts of the large scale structure probes,
 where screening effects take place and therefore reduce the magnitude
of modified gravity effects.  Hence 21cm cosmology offers a prime possibility of observing modification of gravity unhampered by screening effects.  However, it is important to notice that with our specifications of the FFTT, such small scales can be only probed through wavelength modes parallel to the line of sight.

The parametrisation of~\cite{Brax:2011aw}  is used here in the context of generalised chameleon, dilaton and symmetron models, as well as f(R) gravity, which is a special chameleon case.  For all these models (except for dilatons), we find that  21cm observations at reionisation should constrain them more tightly than the large scale structures and CMB observations.  The 21cm signal appears to be also a good discriminator of modified gravity models.

Some models are already tightly
constrained by the tests of gravity in the solar system, in the laboratory and in galactic environments.  If these constraints are imposed, predictions for 21 cm cosmology
are very similar to that of $\Lambda$CDM and signatures of modified gravity should be hardly observable. However, in the case of generalised
symmetron models, strong constraints could be established with 21cm observations.

We have also considered the effect of the scalar field coupling to
photons in modified gravity models. This coupling arises naturally from the
conformal anomaly \cite{Brax:2010uq} and gives rise to a time variation of the
fine-structure constant. Since the 21 cm line is very sensitive to
any variation in the fine structure constant, it can be used to probe the coupling to photons.  This has enabled us to forecast tight bounds on this coupling on the basis of a 21cm experiment covering redshifts going from the beginning to the completion of the reionization.

More accurate predictions will require sophisticated MMCM and Fisher matrix
techniques. This will allow us to consider the whole ensemble of wavelength modes accessible to the experiment and to perform a multi-redshift analysis in order to set precise bounds on the various model parameters, taking account the possible degeneracies with other cosmological and  reionisation parameters.   This is left for future work.

\section{Acknowledgements}
The work of ACD is supported in part by STFC.   S.C. is supported by the Wiener Anspach foundation.

\bibliography{biblio21cm}

\end{document}